\documentclass{ctic}

\usepackage{graphicx} 

\usepackage{graphics} 
\usepackage{epsfig} 

\newtheorem{theorem}{Theorem}[section]

\newtheorem{lemma}[theorem]{Lemma}

\begin{document}

\title{A Note on Gradually Varied Functions and Harmonic Functions }

\author{Li Chen$^{1,4}$, Yong Liu$^{2,1}$, and Feng Luo$^{3,4}$}
\address{$^{1}$~Department of Computer Science \& IT, University of the District of Columbia\\
$^{2}$~Freddie Mac\\
$^{3}$~Department of Mathematics, Rutgers University\\
$^{4}$ Center for Discrete Mathematics and Theoretical Computer Science (DIMACS),\\
       Rutgers University\\
\texttt{lchen@udc.edu}; \texttt{yong\_liu88@yahoo.com}; \texttt{fluo@math.rutgers.edu}
}

\maketitle

\begin {abstract}
Any constructive continuous function must have a gradually varied approximation in compact space. 
However, the refinement of domain for $\sigma-$-net might be very small. Keeping the original 
 discretization (square or triangulation), can we  get some interesting properties related to gradual variation?     
In this note, we try to prove that many harmonic functions are gradually varied or near gradually varied;
 this means that the value of the center point differs from that of its neighbor at most by 2. 
It is obvious that most of the gradually varied functions 
are not harmonic.
This note discusses some of the basic harmonic functions in relation to gradually varied
functions. 

\end {abstract}

\section {Introduction}

In this note, we will discuss some interesting facts about gradually varied functions 
(GVF) and harmonic functions. The compatibility between gradually varied functions 
and harmonic functions is important to the applications of gradually varied functions in real world engineering problems.
   
Any constructive continuous function must have a gradually varied approximation in compact space \cite{Che2005}. 
However, the refinement of domain for $\sigma-$-net might be very small. Keeping the original 
discretization (square or triangulation), we can obtain some interesting properties related to gradual variation.     
In this note, we try to prove that many harmonic functions are gradually varied or near gradually varied,
 meaning that the value of the center point differs from that of its neighbor by at most 2. 
It is obvious that most of the gradually varied functions 
are not harmonic.
This note discusses some of the basic harmonic functions in relation to gradually varied
functions. 
 
Let $A_1, A_2, ..., A_n$ be rational numbers and $A_1< A_2< ...< A_n$. Let $D$ be a graph.
$f: D\rightarrow \{A_1,...,A_n\}$ 
is said to be gradually varied if for any adjacent pair $p,q$ in $D$ and $f(p)=A_{i}$, then
$f(q) = A_{i-1}$, $A_{i}$, or $A_{i+1}$. We usually let $A_i=i$.
 
Extending the concept of gradual variation to the function in continuous space: 
$f: D\rightarrow R$ is gradually varied if $|p-q|\le 1$ then $|f_q-f_p|\le 1$. Or 

\begin{equation}
|f_q-f_p| \le |p-q|. 
\end{equation}

To some extent, gradual variation is the same as the locally Lipschitz condition.
(However, $A_{i}$ may be defined differently.)

On the other hand, a harmonic function satisfies: 

\begin{equation}
  \frac {\partial ^{2} f }{\partial x ^2} + \frac {\partial ^{2} f }{\partial y ^2} =0 
\end{equation}

A main property of the harmonic function is that for a point $p$, $f(p)$ equals 
the average value of all surrounding points of $p$.


If $f$ is harmonic and $p, q$ are two points such that $f(p)<f(q)$ and 
$s$ is a path (curve) from $p$ to $q$, then we. We know 

\begin{equation}
	f_q-f_p = \int_{p,q} \nabla f\cdot d {\bf s}
\end{equation}

If $s$ is a projection of a geodesic curve on $f$, does the gradient $\nabla f$ maintain some of its properties? 
For example, is it a constant or does it have any property relating to gradual variation?

What we would like to prove is that if we define 

\begin{equation}
  f_{mg} (p,q) =\max \{ |\nabla(f)|\} \mbox {on curve $s$ or entire $D$ }
\end{equation}

\noindent should we have

{\bf Observation A:} 
$f_{mg}(p,q)< 2\cdot |(f_q-f_p)|/ length(s)$ when $f$ is harmonic?

Therefore, our purpose is to show that many basic harmonic solutions are at least ``near'' GVF solutions.

\section { Harmonic Functions with Gradual Variation}

Given the value of a set of points in domain $D$, $f: J\rightarrow R$, $J\subset D$,
for 4-adjacency in 2D (grid space), using an interpolating process, we can obtain a GVF solution.\cite{Che2005} 
We also can solve a linear equation using a fast algorithm for a sparse matrix of the harmonic equation based
on

\begin{equation}
    f_{i,j} = \frac{1}{4} (f_{i-1,j}+f_{i+1,j} + f_{i,j-1}+f_{i,j+1})
\end{equation}

or give an initial value for $f$ and then do an iteration. This formula gives a fast solution and 
also gives a definition of discrete harmonic functions~\cite{Lov}.

How we use the GVF algorithm to guarantee a near harmonic solution is 
a problem. We can use the divide-and-conquer method to have an $O(n log n)$ algorithm and then iterate it a few times to get 
a harmonic solution.

Assume $b_1$ and $b_2$ are two points in boundary $J$. $f(b_1) < f(b_2)$ and $s(b_1,b_2)$ is a path from $b_1$ 
to $b_2$. So

\begin{equation}
     \frac{(f(b_2)-f(b_1))}{length(s(b_1,b_2))}
\end{equation}
 
\noindent is the average slope of the curve. We can define
\begin{equation}
    slope (b_1,b_2) =\max \{ \frac{(f(b_2)-f(b_1))}{length(s(b_1,b_2))} | s(b_1,b_2) \mbox{ is a path}\}
\end{equation}

\noindent Therefore, there is a $s(b_1,b_2)$ whose length reaches the minimum. Such a path will be a geodesic curve. 

With the consideration of the maximum ``slope'', the reason for $Observation A$ is 
\begin{equation}
    |\nabla f|\le  (\frac{\partial f}{ \partial x}^2 + \frac{\partial f}{ \partial y}^2)^{1/2} \le ? 2\cdot slope(b1,b2) \le 2.
\end{equation}

\noindent In general,
\begin{equation}
    |\nabla f|\le  (\frac{\partial f}{ \partial x}^k + \frac{\partial f}{ \partial y}^k)^{1/k} \le ? 2\cdot slope(b1,b2) \le 2.
\end{equation}


\noindent where $k>0$. Since $slope\le 1$ based on the condition of gradual variation, we want to show that the  
harmonic solution is nearly gradually varied. Note that the gradual variation condition is similar 
to the Lipschitz condition. 

There are two reasons for using ``2'' in the above formula as the ratio: (1) It is not possible to use ``1,'' and (2) anything less than 2 is almost
gradual variation.

\begin{lemma}
  There are simple cases in discrete space that the harmonic solution reaches difference 1.5. 
\end{lemma}

{\it Proof} Assume that we have five points in grid space in 
direct adjacency: $(i,j), (i-1,j), (i+1,j) , (i,j-1) , (i,j+1)$ and 
$f_{i-1,j}= 1$, $f_{i+1,j} = f_{i,j-1}=f_{i,j+1}=3$ 

We want to know what $f(i,j)$ equals. Using the GVF, we get $f(i,j)=2$ by Definition 1.1. See Fig. 3.1. 
\begin{figure}[h]
	\begin{center}

   \epsfxsize=2in 
   \epsfbox{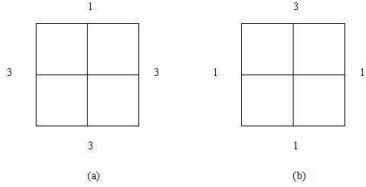}

	\end{center}
\caption{ Discrete harmonic interpolation.}
\end{figure}

Using harmonic functions, we will have $f(i,j)=2.5$ by Definition 2.1. With the same principle,
we can let $f_{i-1,j}= 3$ and $f_{i+1,j} = f_{i,j-1}=f_{i,j+1}=1$. 
So $f(i,j)=1.5$ for the harmonic solution and $f(i,j)=2$ still for gradually varied.  \\

When we use the harmonic solution to approximate gradual variation, we need to see if we can find the best value
when choosing from two possible values. A simple algorithm may be needed to make this decision.

 
{\bf Observation B:} 
There is a GVF that is almost harmonic: $|center-average Of Neighbor|<1$ or $|center-average Of Neighbor|< c$, 
$c$ is a constant.

The above examples show that a perfect GVF is not possible for a harmonic solution. The gradient (maximum directional
derivative) less than $2\cdot f'_{m}$, $f'_{m}$ denotes the maximum average change (slope) of any path between 
two points
on the boundary possessing the mean of gradual variation.  

Every linear function is harmonic. And for quadratic functions, we have

 $f(x,y)= a x^2+b y^2 + c xy $ 
 
\noindent is harmonic if and only if $a=-b$.  However, the following example will not meet the case.

{\bf Example 1} Three vertices of a triangle are $p1=(0,0)$, $p2=(9,0)$, $p3=(-8,4)$. The linear function 
$f(x,y)= x+3y$.   

\begin{figure}[h]
	\begin{center}

   \epsfxsize=2in 
   \epsfbox{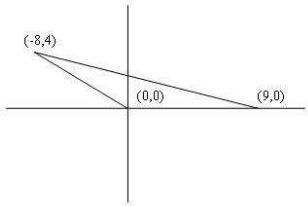}

	\end{center}
\caption{ Example of linear functions.}
\end{figure}

This triangle satisfies the gradually varied conditions:

\[|f(p1)-f(p2)|=9 \le |p1-p2|=9\]

\[|f(p2)-f(p3)|=|9-4| \le |p1-p2|\] 

\[|f(p1)-f(p3)|=|0-4| \le \sqrt{8^2+4^2}\] 

If we consider a point $p=(x,y)$ on the line $<p_2, p_3>$ when $x = 0$ and $y=36/17=2.4$, then $f(x,y) >7$. 
This point and $p_1$ do not 
maintain the condition of gradual variation. $|f(p)-f(p_1)|>7 > |p-p_1|$.\\

This example seems to break the observation we have made. However, let us revisit the function $f(x,y)= x+3y$ and let $z=f(x,y)$. We have $z-x-3y=0$. We can have $y= \frac{1}{3} z- \frac{1}{3} x$ represent the 
triangle and associated function. In general, for a linear function in 3D

\[ a x + b y + c z +d =0 \]

We can always find a coefficient that has the maximum absolute value. We will have the equivalent equation that
has

\begin{equation}
   A X + B Y  + D = Z 
\end{equation}

\noindent where $|A|$ and $|B|\le 1$. This property is often used in computer graphics.\\

\begin{lemma} 
   The Piece-wise linear function preserves the property of gradual variation. 
\end{lemma}

{\it Proof:}  We first want to discuss the case of a single triangle where any piecewise linear function 
is a harmonic function. In this case we can write the function like this

\[ f(x,y) = a x +b y +c , \mbox{\hskip 0.1in} |a|,|b|\le 1 \]

  $\frac{\partial f}{\partial x} = f_x =a$ , $\frac{\partial f}{\partial y} = f_y =b$. 
The gradient is a constant $\sqrt{ a^2+b^2}$. There is a horizontal and vertical 
line that goes through boundary points. The maximum average rate of change  $r$ (average slope on 
the path between two points on the boundary) is greater than or equal to $\max{a,b}$. 

Since $|a|,|b|\le 1$ ; $r \le \sqrt {a^2+b^2}\le \sqrt{2} \max{a,b} \le \sqrt{2}$. 
So $r < 2$. 

If this piecewise linear function is on a polygon (2D), it will still have this property. \\ 

The problem is that in this proof, we have not used the conditions of gradual variation directly.  The conditions are

\[|f(p_1)-f(p_2)|=| a (x1-x2) +b (y1-y2)| \le |p_1 -p_2| =\sqrt {(x1-x2)^2+(y1-y2)^2}\]

\[|f(p_1)-f(p_3)|= |a (x1-x3) +b (y1-y3)| \le   |p_1 -p_3| =   \sqrt {(x1-x3)^2+(y1-y3)^2}\]

\[|f(p_2)-f(p_3)|= |a (x2-x3) +b (y2-y3)| \le  |p_3 -p_2| = \sqrt {(x2-x3)^2+(y2-y3)^2}\]

 We
have used the GVF general property and the triangle constraint. The next section will discuss a more general case.\\

\section {Gradually Varied Semi-Preserving }  

In this section, we extend the content of above sections using more rigorous
mathematical definitions. Harmonic functions can be characterized by the mean value theorem.  
Here we are interested in
 harmonic functions that are gradually varied.  More specifically, a function is said to be
gradually varied semi-preserving if

\begin{equation}
        \max_{D} |\nabla u| \le c \cdot \max_{p,q\in \partial D} \frac{|u(p)-u(q)|}{|p-q|} 
\end{equation}

\noindent where $\nabla u$  is the gradient of $u$, $D$  is a domain with the boundary $\partial D$, 
and $c$  is a constant.

The above formula poses a property of computational importance.  
We can show that linear functions and quadratic hyperbolic functions satisfy 
the condition of gradually varied semi-preserving.

If $u$ is linear we can assume that $u=ax+by+c$ and if $u$ is quadratic hyperbolic 
we can let $u =a (x^2 - y^2)$. We do not restrict the value of $a,b,c$ here.

\begin{lemma}
{\bf Proposition 2}   If $u$  is linear  or quadratic hyperbolic, then 
\begin{equation}
          \max_{B} |\nabla u| \le \sqrt(2) \cdot \max_{p,q\in \partial B} \frac{|u(p)-u(q)|}{|p-q|}                                         
\end{equation}
\noindent where $B$ is any ball.
\end{lemma}

{\it Proof} Let $u$ be a linear function
		$u=ax+by+c$
then  
\begin{equation}
	 |\nabla u| = \sqrt(a^2+b^2) \le \sqrt(2) \max \{|a|,|b|\}  
\end{equation}
	 
On the other hand, if we choose $p=(-r,0), q=(r,0)$ on $\partial B$,

where $r$ is the radius of the ball $B$.  Then 
\begin{equation}
  \frac{|u(p)-u(q)|}{|p-q|}=\frac{|-ar-ar|}{2r} = |a|
\end{equation}
Choosing another pair of $p$ and $q$ on $\partial B$,
	                                           
\[p=(0,r), q=(0,-r)\]

we have
\begin{equation}
	\frac{|u(p)-u(q)|}{|p-q|}=  |b|
\end{equation} 

\noindent Combining (13), (14) and (15) we conclude (12) when  $u$ is linear.

Now, consider $u$ as a quadratic hyperbolic function: $u =a (x^2 - y^2)$.
Then,

\begin{equation}
    |\nabla u| = 2 |a| \sqrt(x^2+y^2) \le 2 |a| r 
\end{equation}

On the other hand, if we choose $p$ and $q$ on $\partial B$, 

\[p=(0,r), q=(r,0)\]

Then

\begin{equation}
   \frac{|u(p)-u(q)|}{|p-q|}=\frac{|-ar^2-ar^2|}{\sqrt(2r^2)} = 2 |a| r 
\end{equation}

\noindent Combining (16) and (17) we have

\[|\nabla u| \le \sqrt(2)\frac{|u(p)-u(q)|}{|p-q|} \le \sqrt(2)\max_{x,y\in \partial B}\frac{|u(x)-u(y)|}{|x-y|}\]

\noindent and (12) follows. \\

\section{Discussion}

Recent studies show an increased interest in connecting discrete mathematics with continuous mathematics, especially
in geometric problems. For instance, the variational principle has been used for triangulated surfaces
in discrete differential geometry, see ~\cite{Luo}.  
This note presented an idea of combining a type of discrete function: the gradually varied function and a type of
 continuous function: the harmonic function in a relatively deep way in terms of continuous mathematics.  
 The harmonic function is a weak solution to
the Dirichet problem which is about how to find a surface when the boundary curve is given.  
The gradually varied function was proposed to solve a filling problem in computer vision.  We are hesitant to use
the method of the  Dirichlet  problem for the discrete filling problem since we do not know the exact formula (function)
on the boundary, even though we know the sample points. This problem is also related to the 
Whitney's problem \cite{Fef}\cite{KZ}. 
Some ideas have been presented by the first author in the Workshop on Whitney's problem in 2009 organized by C. Fefferman and
N. Zobin at College of William and Mary (http://nxzobi.people.wm.edu/whitney/whitney.htm).  
During the Workshop, P. Shvartsman presented an idea of using 
geodesic curves in Sobolev space (See related paper \cite{Shv}). Our idea about using the geodesic curve presented 
in this note was independently obtained.


\end{document}